\documentstyle[eqsecnum,preprint,prd,aps,psfig,epsfig]{revtex}
\tighten
\let\jnfont=\rm
\def\NPB#1,{{\jnfont Nucl.\ Phys.\ }{\bf B#1},}
\def\PLB#1,{{\jnfont Phys.\ Lett.\ B }{\bf #1},}
\def\PRD#1,{{\jnfont Phys.\ Rev.\ D }{\bf #1},}
\def\PRL#1,{{\jnfont Phys.\ Rev.\ Lett.\ }{\bf #1},}
\def\ZPC#1,{{\jnfont Z.~Phys.\ C }{\bf #1},}
\def\ETslash{\not{\hbox{\kern-4pt $E_T$}}}
\begin{document}
\preprint{\parbox{2.0in}{\noindent TU-566 \\ hep-ph/9905486 }}

\title{\ \\[15mm] $R_b$ and $R_{\ell}$ in MSSM without R-Parity}
\vspace{1cm}

\author{\ \\[2mm] Jin Min Yang\footnote{Address after March of 2000:
Institute of Theoretical Physics, Academia Sinica, Beijing, China}} 

\address{ \ \\[2mm]
 {\it Department of Physics, Tohoku University,
               Aoba-ku, Sendai 980-8578, Japan} }
\maketitle
\vspace{1cm}

\begin{abstract}

We examined  $Z\ell^+\ell^-$ and $Zb\bar b$ couplings in 
the minimal supersymmetric model (MSSM) with explicit 
$R$-parity violating interactions. We found the top quark 
L-violating couplings $\lambda'_{i3k}$ and B-violating 
couplings $\lambda''_{3j3}$ could give significant 
contributions through the top quark loops. 
To accomadate the latest $R_{\ell}$ data,
$\lambda'_{i3k}$ are suject to stringent bounds, some of which 
can be much stronger than the current bounds.
Within the current perturbative unitarity bound of 1.25 for
$\lambda''_{3j3}$, the $R_b$ value in $R$-violating MSSM
agrees well with the experimental data at $2\sigma$ level, 
but may lie outside the $1\sigma$ range depending on the involved 
sfermion mass.
\end{abstract}

\section{Introduction}
\label{sec1}
The Standard Model (SM) has been very successful phenomenologically.
Despite its success, the SM is still believed to be a theory effective 
at the electroweak scale and that some new physics must exist at higher 
energy regimes. So far there have been numerous speculations on the 
possible forms of new physics beyond the SM. In the effective Lagrangian 
approach \cite{dim6}, the observable effects of new physics at
the SM energy scale are described in the form of anomalous interactions.
Among the various candidates for new physics which have predictive power
at energies not far above the weak scale, the most intensively studied 
one is the weak-scale minimal supersymmetric model (MSSM)\cite{MSSM}, 
which has many attractive features and is often arguably the most promising 
one.

In the MSSM, the invariance of $R$-parity,  defined by $R=(-1)^{2S+3B+L}$ for 
a field with spin $S$, baryon-number $B$ and  lepton-number $L$,  is often 
imposed on the Lagrangian in order to maintain the separate conservation 
of baryon-number and lepton-number. However, this conservation is not 
dictated by any fundamental principle such as gauge invariance and there 
is no compelling theoretical motivation for it. The most general 
superpotential of the MSSM consistent with the 
$\rm SU(3)\times SU(2)\times U(1)$ symmetry and supersymmetry 
contains $R$-violating interactions which are given by~\cite{rp1}
\begin{equation}\label{poten}
{\cal W}_{\not \! R}=\frac{1}{2}\lambda_{ijk}L_iL_jE_k^c
+\lambda'_{ijk} L_iQ_jD_k^c
+\frac{1}{2}\lambda''_{ijk}\epsilon^{abd}U_{ia}^cD_{jb}^cD_{kd}^c
+\mu_iL_iH_2,
\end{equation}
where $L_i(Q_i)$ and $E_i(U_i,D_i)$ are the left-handed
lepton (quark) doublet and right-handed lepton (quark) singlet chiral 
superfields. $i,j,k$ are generation indices and $c$ denotes charge 
conjugation. $a$, $b$ and $d$ are the color indices and $\epsilon^{abd}$ 
is the total antisymmetric tensor. $H_{1,2}$ are the Higgs-doublets chiral 
superfields. The  $\lambda_{ijk}$ and $\lambda'_{ijk}$ are $L$-violating 
couplings, $\lambda''_{ijk}$ $B$-violating couplings.  
$\lambda_{ijk}$ is antisymmetric in the first two indices and 
$\lambda''_{ijk}$ is antisymmetric in the last two indices.
The phenomenological studies for these $R$-violating couplings were started 
long time ago\cite{rp1}. While this is an interesting problem in its
own right, the recent anomalous events at HERA \cite{HERA} and the evidence
of neutrino oscillation \cite{neutri} might 
provide an additional motivation for the study of these $R$-violating 
couplings. So both theorists and experimentalists have recently intensively 
examined the phenomenology of $R$-parity breaking supersymmetry in various 
processes\cite{rp2,rp3} and obtained some bounds\cite{review}.

It is notable that some of these $R$-violating couplings contribute
to the precisely measured $Zb\bar b$ and $Z\ell^+\ell^-$ couplings 
through sfermion-fermion loops. Since the MSSM is a renormalizable field 
theory and the sparticles get their masses through explicit soft breaking 
terms, the decoupling theorem \cite{dec} implies that the effects of these 
sparticle loops will be suppressed by some orders of $1/M_{SUSY}$ and 
vanish as $M_{SUSY}$ goes far above the weak scale. So, in general, 
these sfermion-fermion loop effects on lower energy observables are small. 
However, due to the non-decoupling property of heavy SM particles 
(which get masses through spontaneous gauge symmetry breaking), the effects 
of the sfermion-fermion loops with the fermion being the top quark 
(hereafter called the top quark loops) may be enhanced by the large top 
quark mass. Therefore, the $R$-violating  top 
quark couplings in Eq.(\ref{poten}), $\lambda'_{i3k}$ and $\lambda''_{3jk}$,
which currently subject to  quite weak bounds (see Ref.~\cite{review} for 
a review), may give rise to significant contributions through the top quark 
loops. 
  
Although some bounds on the $R$-violating couplings were derived 
from $R_{\ell}\equiv \Gamma(Z\to  {\rm hadrons})/\Gamma(Z\to \ell^+ \ell^-)$
a few years ago \cite{R_l}, it is necessary to give a thorough examination 
for the $R$-violating quantum effects on $Zb\bar b$ and $Z\ell^+\ell^-$ 
couplings since the measurements of both
$R_b\equiv \Gamma(Z\to b\bar b)/\Gamma(Z\to {\rm hadrons})$  and $R_{\ell}$ 
have been much improved nowadays \cite{data,mar}. Also, when deriving bounds 
from $R_b$ and $R_{\ell}$, the $R$-conserving MSSM quantum effects, 
which were neglected in previous studies \cite{R_l},  should be included. 
In this paper we study the contributions of the trilinear explicit
$R$-violating interactions to $Zb\bar b$ and $Z\ell^+\ell^-$ couplings.
By using the latest data of  $R_b$ and $R_{\ell}$, we examine the bounds on 
the top quark $R$-violating couplings. The $R$-conserving MSSM 
quantum effects on  $Zb\bar b$ vertex are also taken into account 
in our analyses.

Note that at the level of the superpotential, the explicit $L$-violating terms 
$\mu_i L_iH_2$ can be rotated away by a field redefinition\cite{rp1}. However, 
such a redefinition does not leave the full Lagrangian invariant when 
including the soft-breaking terms\cite{soft}. We in this paper focus on the
trilinear explicit $R$-violating interactions and ignore the effects
of the terms $\mu_i L_iH_2$ or the so-called spontaneous $R$-violation 
induced by the non-zero VEVs of sneutrinos.  

This paper is organized as follows. In Sec. \ref{sec2} 
we calculate the contributions of $R$-violating MSSM to $Zb\bar b$ and  
$Z\ell^+\ell^-$ couplings. In Sec. \ref{sec3} we present the contributions
to $R_b$ and $R_{\ell}$, and derive the limits from the latest experimental
data. Finally, we give the conclusion in  Sec. \ref{sec4}.

\section{$Z b\bar b$ and $Z\ell^+\ell^-$ in $R$-violating MSSM}
\label{sec2}

Neglecting the dipole-moment coupling which are suppressed by $m_b/m_Z$,
the  $R$-violating MSSM contribution to $Zb\bar b$ vertex takes the form 
\begin{eqnarray}\label{form}
\Delta V_{Zb\bar b}=i\frac{e}{s_Wc_W}\gamma_{\mu}\left[P_R g_R^b \Delta_R^b
+P_L g_L^b \Delta_L^b \right],
\end{eqnarray} 
where $P_{R,L}=(1\pm \gamma_5)/2$ and $g_L^b$ ($g_R^b$) is the 
$Zb_L\bar b_L$ ($Zb_R\bar b_R$) coupling in the SM. 
(Throughout this paper the subscripts $R$ and $L$ stand for chirality.)
The new physics contribution factors $\Delta_L^b$ and 
$\Delta_R^b$ comprise of both the $R$-conserving MSSM contribution 
and $R$-violating contribution which are denoted by
\begin{eqnarray}
\Delta_L^b=\Delta_L^{({\rm MSSM})}+\Delta_L^{({\not \! R})},\\
\Delta_R^b=\Delta_R^{({\rm MSSM})}+\Delta_R^{({\not \! R})}.
\end{eqnarray} 

The $R$-conserving MSSM interactions contribute to $R_b$ mainly 
through\cite{zbbsusy}:
\begin{itemize}
\item[{\rm(1)}] Chargino-stop loops. Their contribution is most likely
                sizable since they contain the large $\tilde t_R-b_L-$Higgsino
                Yukawa  coupling squared, which is proportional to 
                $\frac{M_t^2}{M_W^2}(1+\cot^2\beta)$.
\item[{\rm(2)}] Charged and neutral Higgs loops. For a very light CP-odd
                Higgs boson $A^0$ ($50\sim 80$ GeV) and very large 
                $\tan\beta$ ($\sim 50$), their contribution could be sizable
                \cite{zbbsusy}.
\item[{\rm(3)}] Neutralino-sbottom loops. Their contribution is negligibly
                small for low and intermediate $\tan\beta$, but could
                be sizable for very large  $\tan\beta$ \cite{zbbsusy}. 
\end{itemize}
Since the dominant MSSM contribution is from the chargino-stop loops
for low and intermediate $\tan\beta$ ($1\sim 30$), we in our calculation 
consider the chargino-stop loops while give a brief comment
on the effects of other loops. 
A detailed calculation of the full one-loop effects of MSSM on $Zb\bar b$
coupling can be found in \cite{zbbsusy}. Here we present the results 
for the chargino loops. The  Feynman diagrams for  chargino-stop loops
are shown in Fig.1.  The contribution factor 
$\Delta_R^{({\rm MSSM})}\equiv  \Delta_R^{({\tilde t_L})}$ arise 
from the first three diagrams of Fig.1 induced by 
$\tilde t_L-b_R-\tilde \chi^+_j$ Yukawa couplings, while 
$\Delta_L^{({\rm MSSM})}\equiv\Delta_L^{({\tilde t_L})}
+\Delta_L^{({\tilde t_R})}$ with $\Delta_L^{({\tilde t_L})}$ 
arising from the middle three diagrams of Fig.1 induced by 
$\tilde t_L-b_L-\tilde \chi^+_j$ gauge couplings and
$\Delta_L^{({\tilde t_R})}$ from the last three diagrams of Fig.1 
induced by $\tilde t_R-b_L-\tilde \chi^+_j$ Yukawa couplings.
The expressions of $\Delta_R^{(\tilde t_L)}$, $\Delta_L^{({\tilde t_L})}$
and $\Delta_L^{({\tilde t_R})}$ are presented in the Appendix. 

Through loop diagrams, the $R$-violating couplings 
$\lambda''_{ij3}$ and $\lambda'_{ij3}$ 
contribute to $Zb_R\bar b_R$, and  $\lambda'_{i3k}$ 
contribute to  $Zb_L\bar b_L$. (Note that, for example, 
$\lambda''_{ij3}$ can also induce $Zb_L\bar b_L$ coupling 
through loops, which is suppressed by $M_b^2/M_Z^2$ and thus negligibly 
small.)  The Feynman diagrams for the loop contributions of these couplings
to $Zb\bar b$ coupling are shown in Figs.2, 3 and 4, respectively.
Their contributions denoted as $\Delta_R^{(\lambda''_{ij3})}$,
$\Delta_R^{(\lambda'_{ij3})}$, and $\Delta_L^{(\lambda'_{i3k})}$ are 
presented in the Appendix. 

Through loops, the couplings $\lambda_{ijk}$ and $\lambda'_{ijk}$ with 
$i=1,2$ and $3$ contribute to $Z\ell^+\ell^-$ with $\ell=e$,$\mu$ and $\tau$,
respectively.  
The pure leptonic couplings $\lambda_{ijk}$ are not relavent to the 
top quark and will not be considered in our analyses.
The Feynman diagrams of the contribution 
of  $\lambda'_{ijk}$ to $Z\ell^-_L \ell^+_L$ are shown in Fig.5. (The loops 
of  $\lambda'_{ijk}$ can also induce $Z\ell^-_R\ell^+_R$ coupling, 
which is suppressed by $M_{\ell}^2/M_Z^2$ and thus negligibly small.)
The contribution to  $Z\ell^+\ell^-$ vertex takes the form 
\begin{eqnarray}
\Delta V_{Z\ell^+\ell^-}=i\frac{e}{s_Wc_W}\gamma_{\mu}\left[
P_R g_R^e \Delta_R^{\ell}
+P_L g_L^e \Delta_L^{\ell} \right],
\end{eqnarray} 
where $g_L^e$ and $g_R^e$ are the couplings in the SM,  
and $\Delta_L^{\ell}$ and $\Delta_R^{\ell}$ the contributions from
the couplings  $\lambda'_{ijk}$ with $\Delta_R^{\ell}\approx 0$ and 
$\Delta_L^{\ell}$ being obtained from $\Delta_L^{(\lambda'_{i3k})}$ 
with substitutions of $b\to \ell$, $\nu^i\to u^j$ and  
$\tilde\nu^i\to \tilde u^j$.   

We would like to make a few comments on the above calculations:
\begin{itemize}
\item[{\rm(a)}] In our calculations, we used dimensional regularization to 
control the ultraviolet divergences in the virtual loop corrections and 
we adopted the on-mass-shell renormalization scheme.
The ultraviolet divergences in the self-energy and the vertex loops 
are contained in Feynman integrals. We have checked that in our 
results, the ultraviolet divergences cancelled as a result of 
renormalizability of the MSSM. 
\item[{\rm(b)}]
In our calculations, various sfermion states are involved.
We note that in general there exist the mixing between left- and right-handed 
sfermions of each flavor (denoted as $\tilde f_L$ and $\tilde f_R$), 
as suggested by low-energy supergravity models \cite{mix}. 
(We do not consider the flavor mixing of sfermions.)  
So $\tilde f_L$ and $\tilde f_R$ are in general not the physical
states (mass eigenstates), instead they are related to mass eigenstates
 $\tilde f_1$ and $\tilde f_2$ by a unitary rotation:
\begin{eqnarray}
\tilde f_R&=&  \cos\theta \tilde f_1-\sin\theta \tilde f_2,\\
\tilde f_L&=&  \sin\theta \tilde f_1+\cos\theta \tilde f_2.
\end{eqnarray}
In our calculations, we give the results 
in terms of $\tilde f_L$ and $\tilde f_R$, which can be easily applied to
the general case by using the above relations between $\tilde f_{L,R}$
and  $\tilde f_{1,2}$.   
\item[{\rm(c)}] We neglected the $R$-conserving MSSM contribution to 
$Z\ell^+\ell^-$ couplings because they are expected to be small,  unlike the
$Zb \bar b$ case where chargino-stop loops could contribute significantly. 
\item[{\rm(d)}] While it is theoretically possible to have both $B$-violating 
and $L$-violating terms in the Lagrangian, the non-observation of proton decay 
imposes very stringent conditions on their simultaneous presence. In our 
calculation (and in the following numerical calculations) we consider the 
presence of one non-zero coupling at one time. 
\end{itemize}

 Let us neglect the mixing between the left- and right-handed sfermions
of each flavor, and assume the value of 100 GeV for all sparticles and 
find out which coupling could give large contribution.
(For heavier sparticles the contribution becomes smaller.)
The top quark mass is fixed to $175$ GeV throughout the paper. 
The results are found to be
\begin{eqnarray}\label{result1}
{\rm Fig.2:~~~~} \Delta_R^b/ |\lambda''_{ij3}|^2&=&
\left\{ \begin{array}{rl}  -5.62\% & ~~~{\rm for~}  
\lambda''_{3j3}\\
 0.183\% & ~~~{\rm for~}  
\lambda''_{1j3},\lambda''_{2j3}.
\end{array}\right. \\
\label{result2}
{\rm Fig.3:~~~~} \Delta_R^b/ |\lambda'_{ij3}|^2 &=&
\left\{ \begin{array}{rl}  -2.76\% & ~~~{\rm for~}  
\lambda'_{i33}\\
 0.184\% & ~~~{\rm for~}  
\lambda'_{i13}, \lambda'_{i23},
\end{array}\right. \\
\label{result3}
{\rm Fig.4:~~~~} \Delta_L^b/ |\lambda'_{i3k}|^2 &=& 0.09\% \\
{\rm Fig.5:~~~~} \Delta_L^{\ell}/ |\lambda'_{ijk}|^2 &=&
\left\{ \begin{array}{rl}  -0.77\% & ~~~{\rm for~}  
\lambda'_{i3k}\\
 0.09\% & ~~~{\rm for~}  
\lambda'_{i1k}, \lambda'_{i2k},
\end{array}\right. 
\end{eqnarray}
As expected, even for the same magnitudes of the coupling strength, 
the top quark interactions contribute more significantly than others,
i.e.,  $\lambda''_{3j3}$ and 
$\lambda'_{i33}$ contribute more significantly to $Zb_R\bar b_R$, while
$\lambda'_{i3k}$ to $Z\ell^-_L\ell^+_L$.
In each case, the dominant contribution is found to arise from the top 
quark loops.  

As for the contribution from $R$-conserving MSSM, for comparison, we assume 
$\theta_t=0$ (i.e., $M_{\tilde t_R}=M_{\tilde t_1}$ and  
$M_{\tilde t_L}=M_{\tilde t_2}$), and the masses of both stops
take the value of 100 GeV. Assuming $M=250$ GeV and $\mu=-100$ GeV,
then for $\tan\beta=1~(30)$, the contributions are found to be
\begin{eqnarray}\label{MSSM1}
\Delta_R^{(\tilde t_L)} &=& -0.001\%~ ( -0.28\%)\\ \label{MSSM2}
\Delta_L^{(\tilde t_L)} &=& -0.03\% ~ (-0.02\% ) \\ \label{MSSM3}
\Delta_L^{(\tilde t_R)} &=& 0.27\%  ~ ( 0.17\% )
\end{eqnarray}
As expected, $\Delta_L^{(\tilde t_R)}$ is large because
it is proportional to $\frac{M_t^2}{M_W^2}(1+\cot^2\beta)$.
For large  $\tan\beta$,  $\Delta_R^{(\tilde t_L)}$
also becomes sizable because it is  enhanced by
the factor $\frac{M_b^2}{M_W^2}(1+\tan^2\beta)$.
$\Delta_L^{(\tilde t_L)}$ arise from the gauge coupling and is
found to be always small.
We realize that although the magnitude of $\Delta_R^{(\tilde t_L)}$ 
becomes comparable to that of  $\Delta_L^{(\tilde t_R)}$ 
for large  $\tan\beta$, its contribution to $R_b$ is suppressed by the 
factor $(g^b_R/g^b_L)^2\approx 1/30$ relative to the contribution of 
$\Delta_L^{(\tilde t_R)}$. So, for low and intermediate $\tan\beta$, 
the MSSM contribution is dominated by the last three diagrams of Fig.1, 
which are induced by the $\tilde t_R-b_L-$Higgsino Yukawa coupling. 
Since the sign of $\Delta_R^{(\tilde t_L)}$ is opposite to that of
$\Delta_L^{(\tilde t_R)}$, the MSSM contribution to $R_b$ is relatively 
large when $\tilde t_R$ is the lighter stop $\tilde t_1$ ($\theta_t=0$), 
$\tan\beta$ takes the small value and the lighter chargino is 
Higgsino-like.

Comparing Eqs.(\ref{result1},\ref{result2}) with Eqs.(\ref{MSSM3}), we find 
that for $|\lambda'|\approx 1$ or  $|\lambda''|\approx 1$
the $R$-violating contribution to $R_b$ is of comparable magnitude 
to the MSSM contribution. (Here again we note the fact 
that the contribution of  $\Delta_R^b$ to $R_b$ is suppressed by the 
factor $(g^b_R/g^b_L)^2\approx 1/30$ relative to the contribution of 
$\Delta_L^b$.)  If the stops are significantly
lighter than other sfermions which appear in the top quark loops of 
$R$-violating contributions, the MSSM contribution to $R_b$ can 
be more sizable. So the MSSM contributions should be considered
when deriving the bounds on $R$-violating couplings from $R_b$.  

\section{$R_b$ and $R_{\ell}$ in $R$-violating MSSM}
\label{sec3}

From the results of the preceding section, we found that the 
contributions of $R$-violating top quark interactions 
to both $Zb\bar b$ and  $Z\ell^+\ell^-$ could be significant, 
with the dominant contributions arising from the top quark loops.
The magnitudes of the contributions are proportional to
the relevant coupling strenth squared. 

For the $L$-violating top quark couplings $\lambda'_{i3k}$,
only $\lambda'_{131}$, $\lambda'_{231}$ and  $\lambda'_{133}$ 
are already strongly constrained by atomic parity violation, 
$\nu_{\mu}$ deep-inelastic scattering and $\nu_e$-mass~\cite{review}, 
respectively.  In the case of  the $B$-violating top quark 
couplings $\lambda''_{3j3}$, none of them have been well constrained by 
other processes. Some theoretical bounds on  $\lambda''_{3j3}$
can be derived under specific assumptions.
The constraint of perturbative unitarity at the SUSY breaking scale 
$M_{SUSY}$  would require all the couplings 
$|\lambda''|^2/(4\pi)<1$, i.e., 
$|\lambda''|<3.54$.  
A stronger bound can be obtained if we 
assume the gauge group unification at $M_U=2\times 10^{16}$ GeV and the 
Yukawa couplings $Y_t, Y_b$ and $Y_{\tau}$ to remain in the perturbative 
domain in the whole range up to $M_U$. They imply $Y_i(\mu)<1$ for 
$\mu<2\times 10^{16}$ GeV. Then we obtain an upper bound of 1.25 
for all $\lambda''$ \cite{uni}. 
So it is likely for these top quark $R$-violating couplings, subject
to the current limits, cause large effects on  $R_b$ and/or $R_{\ell}$
and, as a result, their current bounds could be improved further.

In the following we present some representative results. 
For simplicity, we again neglect the mixing between the left- and 
right-handed sfermion states for each flavor, and further assume the 
mass degeneracy for all sfermions. (We will comment on the mixing
effects of stops or sbottoms later.)   
For the parameters $M$ and $\mu$ in the chargino sector,
we take two  representative scenarios: $M=250$ GeV and $\mu=-100$ GeV
(Higgsino-like), and $M=100$ GeV and $\mu=-250$ GeV (gaugino-like). 
In the above two scenarios, the lighter chargino mass
is  112 GeV for  $\tan\beta=2$ and  92 GeV for  $\tan\beta=30$. 
(Note that the lower bound of 91 GeV on  the chargino mass was 
obtained from the LEP runs at c.m. energy of 183 GeV\cite{ALEPH}.) 

\subsection{  $R_b$ in MSSM with $\lambda''_{3j3}$ } 

The coupling $\lambda''_{3j3}$ ($j=1$ or $2$) 
contributes to both $Z\to b\bar b$ though the $t-\tilde d^j_R$
loops and $Z\to d^j\bar d^j$ through $t-\tilde b_R$ loops.
If we assume the mass degeneracy between $\tilde b_R$
and $\tilde d^j_R$, and neglect the masses of the final quark
states, then the contribution to $R_b$ is given by
\begin{eqnarray}\label{R_b}
\Delta R_b&=&2 (1-\xi R_b^{SM})
     R_b^{SM} \frac{\Delta_L^b+(g_R^b/g_L^b)^2 \Delta_R^b}{1+(g_R^b/g_L^b)^2},
\end{eqnarray}
where $\xi= 2$, $\Delta_L^b\approx 0$ and 
$\Delta_R^b=\Delta_R^{(\lambda''_{3j3})}$ given in the
Appendix. The MSSM contribution to  $R_b$ is given by Eq.(\ref{R_b}) 
with  $\xi= 1$ since only $Z\to b\bar b$  in the
hadronic decays of $Z$ could get sizable contribution. 

 We found $\Delta R_b^{({\rm MSSM})}$ is positive and
$\Delta R_b^{(\lambda''_{3j3})}$ is negative.
The combined contribution $\Delta R_b$ is negative for 
$\lambda''_{3j3}$ being of ${\cal O}(1)$.
In Fig.6 we fix $\lambda''_{3j3}=1.25$ and
plot $-\Delta R_b$ as a function of sfermion mass.
The limits shown in Fig.6 are from 
the experimental data and the SM value \cite{data}
\begin{eqnarray}
R_b^{exp}=0.21642\pm0.00073,~~R_b^{SM}=0.2158\pm 0.0002.
\end{eqnarray}
The magnitude of $\Delta R_b$ in scenario A (Higgsino-like) is 
reletively small because this scenario gives the reletively large 
destructive contribution $\Delta R_b^{({\rm MSSM})}$.   
As shown in Fig.6, $\Delta R_b$ lies within the $2\sigma$ range, 
but goes outside the $1\sigma$ range for light sfermion mass.
So the perturbative unitarity bound of 1.25 on 
$\lambda''_{3j3}$ cannot be improved
at  $2\sigma$ level, but can be improved at  $1\sigma$ level.
The limits on $\lambda''_{3j3}$ are listed in 
Table  \ref{table_1}.

Note that the coupling $\lambda'_{i33}$ only contributes to $Z\to b\bar b$, 
and, therefore, the corresponding contribution to $R_b$ is given by 
Eq.(\ref{R_b}) with  $\xi= 1$, and $\Delta_L^b=\Delta_L^{(\lambda'_{i33})}$ 
and $\Delta_R^b=\Delta_R^{(\lambda'_{i33})}$ given in the
Appendix. In this case, there exist both $\Delta_R^b$,
induced dominantly from $t-\tilde e^i_L$ loops in Fig.3, and $\Delta_L^b$,
induced from $b-\tilde \nu^i_L$ and $\tilde b_R-\nu^i$ loops in Fig.4.
As shown in Eqs.(\ref{result2},\ref{result3}), if the masses 
of $\tilde e^i_L$, $\tilde \nu^i_L$ and $\tilde b_R$
are the same (or at least not very different),     
$\Delta_L^b$ is relatively small compared with  
$\Delta_R^b$ because the loops in Fig.4 do not involve the 
top quark. However, the contribution of  $\Delta_L^b$
to $R_b$ is of comparable magnitude to that of  $\Delta_R^b$
due to the suppression factor $(g^b_R/g^b_L)^2\approx 1/30$ for
the latter. Since the two contributions have the opposite sign, they cancel
to a large extent and, therefore, lead to negligibly small contribution to 
$R_b$.
               
\subsection{  $R_{\ell}$ in MSSM with $\lambda''_{3j3}$}
 
The MSSM corrections contribute to $R_{\ell}$ through
their effects on  $Z\to b\bar b$, which is given by
\begin{eqnarray}\label{R_l}
\Delta R_{\ell}&=&\xi R_b^{SM} R_{\ell}^{SM} 
\frac{\Delta_L^b+(g_R^b/g_L^b)^2 \Delta_R^b}{1+(g_R^b/g_L^b)^2},
\end{eqnarray}
with $\xi=2$, $\Delta_L^b=\Delta_L^{(\tilde t_L)}+\Delta_L^{(\tilde t_R)}$
and   $\Delta_R^b=\Delta_R^{(\tilde t_L)}$ given in the Appendix.

The coupling $\lambda''_{3j3}$ contributes to 
$R_{\ell}$ through its effects on  $Z\to b\bar b$ and  $Z\to d^j\bar d^j$,
which is given by Eq.(\ref{R_l}) with  $\xi=4$ under the assumption that 
sfermions involved in the relevant loops have the same mass. 
( Note that under the same assumption,  
the bounds on $\lambda''_{3j3}$ from  $R_{\ell}$ also apply
to $\lambda''_{3jk}$ which contributes to
$Z\to d^k\bar d^k$ and  $Z\to d^j\bar d^j$.) 
Since $\lambda''_{3j3}$  do not directly couple to 
any leptonic flavor, we assume leptonic universality in $R_{\ell}$
in this case.

 We found $\Delta R_{\ell}^{({\rm MSSM})}$ is positive and
$\Delta R_{\ell}^{(\lambda''_{3j3})}$ is negative.
For $\lambda''_{3j3}$ being of ${\cal O}(1)$, 
the combined contribution $\Delta R_{\ell}$ is negative.
In Fig.7 we fix $\lambda''_{3j3}=1.25$ and
plot $-\Delta R_{\ell}$ as a function of sfermion mass.
The upper limits in Fig.7 are obtained from 
the experimental data \cite{data}
\begin{eqnarray}
R_{\ell}^{\rm exp}=20.768\pm 0.024,~~R_{\ell}^{\rm SM}=20.748. 
\end{eqnarray}
From Fig.7 we see that if $\lambda''_{3j3}$ takes
the largest value allowed by perturbative unitarity, the magnitude
of  $\Delta R_{\ell}$ lies outside the $1\sigma$ range for sfermion
mass less than 1 TeV and outside the $2\sigma$ range for   
sfermion mass less than about 200 GeV. So the  perturbative unitarity
bounds on $\lambda''_{3j3}$ can be improved, as 
listed in Table \ref{table_2}.

\subsection{$R_{\ell}$ in MSSM with $\lambda'_{i3k}$} 

The coupling $\lambda'_{i33}$ contributes to $R_{\ell_i}$,  
($\ell_i=e, \mu$ and $\tau$ for $i=1,2$ and $3$, respectively)
through their effects on  $Z\to \ell^-_i\ell^+_i$ and  $Z\to b\bar b$,
which is given by
\begin{eqnarray}
\Delta R_{\ell_i}&=&2 R_{\ell_i}^{SM}\left[ R_b^{SM} 
\frac{\Delta_L^b+(g_R^b/g_L^b)^2 \Delta_R^b}{1+(g_R^b/g_L^b)^2}
-\frac{\Delta_L^{\ell_i}+(g_R^e/g_L^e)^2 \Delta_R^{\ell_i}}{1+(g_R^e/g_L^e)^2}
\right ].
\end{eqnarray}                  
Under the assumption that sfermions involved have the same mass,
the effects of $\lambda'_{i33}$ is the same as $\lambda'_{i3k}$  
which contributes to $Z\to \ell^-_i\ell^+_i$ and  $Z\to d^k\bar d^k$.
The $R$-conserving MSSM contribution to each $R_{\ell_i}$ can be obtained 
from Eq.(\ref{R_l}) by the obvious substitution $\ell \to \ell_i$.

We found both $\Delta R_{\ell_i}^{({\rm MSSM})}$ 
and  $\Delta R_{\ell_i}^{(\lambda'_{i3k})}$ 
are positive. For sfermion mass of 200 GeV, the combined 
contrtribution $\Delta R_{\ell_i}$ versus $\lambda'_{i3k}$ 
is ploted in Figs.8, 9 and 10 for $i=1, 2$ and $3$, respectively.  
The magnitude of each $\Delta R_{\ell_i}$ in scenario A (Higgsino-like) is 
reletively large because this scenario gives the reletively large 
constructive contribution $\Delta R_{\ell_i}^{({\rm MSSM})}$.
The limits ploted in  Figs.8-10 are obtained from
experimental data and the SM values \cite{data} 
\begin{eqnarray}
R_{e}^{\rm exp}=20.803 \pm 0.049,~~ R_{e}^{\rm SM}=20.748\pm 0.019,\\ 
R_{\mu}^{\rm exp}=20.786 \pm 0.033,~~ R_{\mu}^{\rm SM}=20.749\pm 0.019,\\ 
R_{\tau}^{\rm exp}=20.764 \pm 0.045,~~ R_{\tau}^{\rm SM}=20.794\pm 0.019.
\end{eqnarray}                  
From  Figs.8-10 we see that the contribution in each case may
lie outside the $2\sigma$ range, depending on the coupling strength.
The limits on  $\lambda''_{13k}$,  $\lambda''_{23k}$
and  $\lambda''_{33k}$ are listed in Tables 
\ref{table_3}, \ref{table_4} and \ref{table_5}, respectively.

\section{Discussions and Conclusion}
\label{sec4}         

A few remarks are due regarding the numerical results:
\begin{itemize}
\item[{\rm(1)}] The $R$-conserving MSSM effects are not negligible in 
deriving the bounds on the $R$-violating couplings. For example, for sfermion 
mass of 100 GeV, the $1\sigma$ bounds with (without) the MSSM effects are
\begin{eqnarray}
\vert \lambda''_{3j3}\vert < 1.07 (0.55)~~~~({\rm from~} R_b),\\
\vert \lambda''_{3j3}\vert < 0.65 (0.35)~~~~({\rm from~} R_{\ell}),\\
\vert \lambda'_{13k}\vert < 0.73 (0.77)~~~~({\rm from~} R_e),\\
\vert \lambda'_{23k}\vert < 0.60 (0.64)~~~~({\rm from~} R_{\mu}),\\
\vert \lambda'_{33k}\vert < 0.22 (0.32)~~~~({\rm from~} R_{\tau}),
\end{eqnarray}
where $\tan\beta=2$, and $M=100$ GeV and $\mu=-250$ GeV.

\item[{\rm(2)}] 
We note that the limits are not very sensitive to $\tan\beta$
in the range of $\tan\beta>1$. This is because the $R$-conserving
MSSM effects are dominated by the $\tilde t_R-b_L-$Higgsino Yukawa 
coupling squared $\sim\frac{M_t^2}{M_W^2}(1+\cot^2\beta)$ which 
is not sensitive to $\tan\beta$ in the range of intermediate and large 
$\tan\beta$. (Of course, the MSSM effects also have a mild dependence 
on $\tan\beta$ through chargino masses and the unitary matrice diagonalising
the chargino mass matrix.)
It is obvious that as $\tan\beta$ goes lower than 1 
(which is disfavored by the existing experimental data), the MSSM effects 
will be greatly enhanced and thus the results will be very sensitive to 
$\tan\beta$. 

\item[{\rm(3)}] For the MSSM contribution to $Z b\bar b$ vertex, we only
considered the most important part, i.e., the chargino-stop loops.
Since the chargino-stop loops contain the large $\tilde t_R-b_L-$Higgsino 
Yukawa coupling squared, which is proportional to 
$\frac{M_t^2}{M_W^2}(1+\cot^2\beta)$, they are the dominant MSSM effects
in a large part of SUSY parameter space, typically with low or intermediate 
$\tan\beta$ ($1\sim 30$).  For a very light CP-odd Higgs boson $A^0$ 
($50\sim 80$ GeV) and very large $\tan\beta$ ($\sim 50$), the Higgs loops may 
give rise to sizable effects.  For very large $\tan\beta$ and light 
sbottoms, the contribution from the neutralino-sbottom loops may also 
be sizable. In both cases, the contributions to $R_b$ are
positive (adds to the effects of chargino-stop loops), and  
therefore, the bounds will get weaker for $\lambda''_{3j3}$ and 
stronger for $\lambda'_{i3k}$. 

\item[{\rm(4)}] We would like to elaborate again on the effects of
sfermion mixings.
In our numerical calculations, we neglected the mixing between the left- 
and right-handed sfermion states for each flavor, and further assume the 
mass degeneracy for all sfermions. This might be a good approximation
for all sfermions except stops and sbottoms, because the mixing
is proportional to the corresponding fermion mass\cite{mix}. The non-diagonal
element is give by\cite{mix}  $M_{LR}=m_t(\mu\cot\beta+A_t)$ for stop mass 
matrix and  $M_{LR}=m_b(\mu\tan\beta+A_b)$, with $A_t$ and $A_b$ being the 
coefficients of the trilinear soft SUSY-breaking terms
$\tilde t_L\tilde t_R H_2$ and $\tilde b_L\tilde b_R H_1$, respectively.
So in general the stop mixing is significant and sbottom mixing 
can also be significant for large $\tan\beta$. This implies that
the lighter stop ($\tilde t_1$) or/and the lighter sbottom ($\tilde b_1$)
could be significantly lighter than other sfermions
\footnote{For the lighter stop, the direct search from all four experiments at 
LEP give a lower mass bound of 75 GeV \cite{LEP}.
The D0 collaboration at FNAL searched for the jets plus 
${\large \not} \! E_T$ signal of stop and obtained the lower
mass limit of 90 GeV \cite{D0}. However, we should note these bounds may not 
be applicable to the $R$-violating MSSM since the LSP is no longer stable and 
thus the basic SUSY signal is no longer the missing energy.}.  
The stop $\tilde t_1$ is involved in Fig.1 
and the sbottom  $\tilde b_1$ is involved in Fig.5 with $k=3$. 
So if the lighter stop is significantly lighter than other sfermions, 
the MSSM contributions from Fig.1 will be more significant and thus
the limits get weaker for $\lambda''_{3j3}$ and stronger for 
$\lambda'_{i3k}$; if the lighter sbottom is significantly 
lighter than other sfermions (e.g., in case of very large $\tan\beta$),
the couplings $\lambda'_{i3k}$ will have larger effects on $Z\ell^+\ell^-$ 
vertex and thus subject to even stronger bounds. The numerical results 
we presented correspond to the special case $A_t=-\mu\cot\beta$ 
($\theta_t=0$) and $A_b=-\mu\tan\beta$ ($\theta_b=0$).

\item[{\rm(5)}] In our numerical calculation we worked in the  general MSSM,
where  the SUSY parameters are arbitrary at the weak scale and thus we have 
the full parameter space freedom. We note that there are some other popular 
frameworks called the constrained MSSM models, in which the MSSM is usually
embedded in some grand unification scenarios. In such frameworks, there are
only a few free parameters at the grand unification scale and all the
parameters at the weak scale are  generated through the renormalization 
group equations. We did not consider such models in our analyses.

\item[{\rm(6)}]
In our analyses we evaluated $R_b$ and did not present the calculation 
for $b\bar b$ forward-backward asymmetry $A_b$. 
The experimental value of $A_b$ shows a $2.7\sigma$ deviation from the 
SM prediction\cite{data}. To accomadate both  $A_b$ and $R_b$ data,
the new physics contribution has to shift the left- and right-handed
$Zb\bar b$ couplings by  $\sim -1\%$ and  $\sim 30\%$, respectively\cite{mar}.
While the contribution of  $-1\%$  might easily be interpreted as a new 
physics quantum loop correction, as shown by our calculation in $R$-violating
MSSM, a large shift of $30\%$ for right-handed coupling seems too 
strange to explain.  For this reason, it is believed the anomaly of 
$A_b$ stems from a statistical or systematic effect.
From our results we see that although the couplings 
$\lambda''_{ij3}$ and $\lambda'_{ij3}$ contribute to right-handed 
$Zb\bar b$ coupling, they cannot provide an explanation for $A_b$ 
because their contributions are negative. 
\end{itemize}
 
In summary, we evaluated the quantum effects of the trilinear 
$R$-parity violating interactions on $Z\ell^+\ell^-$ and $Zb\bar b$ couplings 
in the MSSM. We found the top quark $R$-violating couplings could give 
significant contributions 
to  $Z\ell^+\ell^-$ and  $Zb\bar b$ through the top quark loops. 
We calculated (1) $R_b$ value in the MSSM with $\lambda''_{3j3}$;
(2)  $R_{\ell}$  value in the MSSM with $\lambda''_{3j3}$;
(3)  $R_{\ell_i}$ values in the  MSSM with $\lambda'_{i3k}$ ($i=1,2,3$).
We found that within the current perturbative unitarity bound of 1.25 for
$\lambda''_{3j3}$, the $R_b$ value agrees well with the experimental data 
at $2\sigma$ level, but may lie outside the $1\sigma$ range for light
the involved sfermion mass. The $R_{\ell}$ data constrained $\lambda''_{3j3}$
more severely than  $R_b$ data. To accomadate the $R_{\ell}$ 
($\ell=e,\mu,\tau$) data, $\lambda'_{i3k}$ are suject to stringent bounds.
A summary of updated current bounds on all $R$-violating top quark Yukawa 
couplings are given in Table \ref{table_6} for slepton mass and squark mass 
of 100 GeV. (Here the squark mass of 100 GeV is chosen just for 
illustration and for the convinienece of comparison with other bounds.) 
Since the bounds from $R_{\ell}$ depend slightly on other SUSY parameters 
of the MSSM, we took the most conservative bounds from our results.

\section*{Acknowledgments}
The author thanks Ken-ichi Hikasa for useful discussions.
The work is supported in part by the 
Grant-in-Aid for Scientific Research (No.~10640243) and Grant-in-Aid 
for JSPS Fellows (No.~97317) from the Japan Ministry of Education, 
Science, Sports, and Culture.  

\section*{Appendix}

 The contribution factor $\Delta_R^{(\tilde t_L)}$ arise from the first
three diagrams of Fig1, $\Delta_L^{(\tilde t_L)}$ from  
the middle three  diagrams of Fig.1 and  $\Delta_L^{(\tilde t_R)}$ 
from  the last three diagrams of Fig.1, which are given by
\begin{eqnarray}
\Delta_R^{(\tilde t_L)}&=&-\frac{g^2}{16\pi^2}
\left (\frac{M_b}{\sqrt 2 M_W \cos\beta}\right )^2
\left\{
-|U_{j2}|^2 B_1(M_b,M_{\tilde \chi_j},M_{\tilde t_L})  \right.\nonumber\\
& &  +U_{j2}^*U_{i2}\frac{O'^R_{ij}}{g^b_R}
    \left [0.5-2C_{24}-M_Z^2(C_{11}-C_{12}+C_{21}-C_{23}) \right.\nonumber\\
& &  \left.
    +\frac{O'^L_{ij}}{O'^R_{ij}}M^2_{\tilde \chi_j}C_0 \right] 
    (k,-p_{\bar b},M_{\tilde \chi_j},M_{\tilde \chi_j},M_{\tilde t_L})
   \nonumber\\
& &  \left. -\frac{g^t_L}{g^b_R}|U_{j2}|^2 2C_{24}
    (p_b,-k,M_{\tilde \chi_j},M_{\tilde t_L},M_{\tilde t_L})\right\},\\
\Delta_L^{(\tilde t_L)}&=&-\frac{g^2}{16\pi^2}\left\{
-|V_{j1}|^2 B_1(M_b,M_{\tilde \chi_j},M_{\tilde t_L})  \right.\nonumber\\
& &  +V_{j1}V_{i1}^*\frac{O'^L_{ij}}{g^b_L}
    \left [0.5-2C_{24}-M_Z^2(C_{11}-C_{12}+C_{21}-C_{23}) \right.\nonumber\\
& &  \left.
    +\frac{O'^R_{ij}}{O'^L_{ij}}M^2_{\tilde \chi_j}C_0 \right] 
    (k,-p_{\bar b},M_{\tilde \chi_j},M_{\tilde \chi_j},M_{\tilde t_L})
   \nonumber\\
& &  \left. -\frac{g^t_L}{g^b_R}|V_{j1}|^2 2C_{24}
    (p_b,-k,M_{\tilde \chi_j},M_{\tilde t_L},M_{\tilde t_L})\right\},\\
\Delta_L^{(\tilde t_R)}&=&-\frac{g^2}{16\pi^2}
    \left (\frac{M_t}{\sqrt 2 M_W \sin\beta}\right )^2\left\{
-|V_{j2}|^2 B_1(M_b,M_{\tilde \chi_j},M_{\tilde t_R})  \right.\nonumber\\
& &  +V_{j2}V_{i2}^*\frac{O'^L_{ij}}{g^b_L}
    \left [0.5-2C_{24}-M_Z^2(C_{11}-C_{12}+C_{21}-C_{23}) \right.\nonumber\\
& &  \left.
    +\frac{O'^R_{ij}}{O'^L_{ij}}M^2_{\tilde \chi_j}C_0 \right] 
    (k,-p_{\bar b},M_{\tilde \chi_j},M_{\tilde \chi_j},M_{\tilde t_R})
   \nonumber\\
& &  \left. -\frac{g^t_R}{g^b_R}|V_{j2}|^2 2C_{24}
    (p_b,-k,M_{\tilde \chi_j},M_{\tilde t_L},M_{\tilde t_R})\right\}.
\end{eqnarray} 
Here the functions $B_1$ and $C_{ij}$, $C_0$ are 2- and 3-point 
Feynman integrals defined in \cite{func}, and their functional dependences 
are indicated in the bracket following them with $k$, $p_b$ and  $p_{\bar b}$ 
being the momentum of $Z$-boson, $b$ and $\bar b$, respectively.
The $O'^L_{ij}$ and  $O'^R_{ij}$ are defined by
$O'^L_{ij}=-V_{i1}V_{j1}^*-V_{i2}V_{j2}^*/2+\delta_{ij}
\sin^2\theta_W$ and $O'^R_{ij}=-U_{i1}^*U{j1}-U_{i2}^*U_{j2}/2+\delta_{ij}
\sin^2\theta_W$, respectively. 
The unitary matrix elements $U_{ij}$ and $V_{ij}$, and
the chargino masses $\tilde M_j$ depend on the parameters $M$, $\mu$ and
 $\tan\beta$ via Eq.(c18)-(c21) of  Ref.\cite{MSSM}.
Here we defined $\tan\beta=v_2/v_1$ with $v_2$ ($v_1$) being the vev
of the Higgs doublet giving up-type (down-type) quark masses,
so $\theta_v$ in \cite{MSSM} should be substituted by $\pi/2-\beta$.
$M$ is the  $SU(2)$ gaugino masses and $\mu$ is the coefficient of
the $H_1H_2$ mixing term in the superpotential.

The contribution of Fig.2 to  $Zb_R\bar b_R$ coupling is given by 
\begin{eqnarray}\label{deltR1}
\Delta_R^{(\lambda''_{ij3})}&=&
-|\lambda''_{ij3}|^2 
\frac{f_c}{16\pi^2}
\left\{ -B_1(M_b,M_{d^j},M_{\tilde u^i_R})-B_1(M_b,M_{u^i},M_{\tilde d^j_R})
                                    \right.\nonumber\\
& & +2\frac{g^{\tilde u^i_R}_R}{g^b_R}
C_{24}(p_b,-k,M_{d^j},M_{\tilde u^i_R},M_{\tilde u^i_R})
+2\frac{g^{\tilde d^j_R}_R}{g^b_R}
C_{24}(p_b,-k,M_{u^i},M_{\tilde d^j_R},M_{\tilde d^j_R})\nonumber\\
& & -\frac{g^{d^j}_R}{g^b_R}\left[0.5-2C_{24}-
M_Z^2(C_{11}-C_{12}+C_{21}-C_{23}) \right.\nonumber\\
& &  \left.+\frac{g^{d^j}_L}{g^{d^j}_R}M_{d^j}^2
C_0 \right] (k,-p_{\bar b},M_{d^j},M_{d^j},M_{\tilde u^i_R})\nonumber\\
& &   -\frac{g^{u^i}_R}{g^b_R}\left[0.5-2C_{24}-
M_Z^2(C_{11}-C_{12}+C_{21}-C_{23}) \right.\nonumber\\
& &  \left.\left.+\frac{g^{u^i}_L}{g^{u^i}_R}M_{u^i}^2
C_0 \right ] (k,-p_{\bar b},M_{u^i},M_{u^i},M_{\tilde d^j_R})\right\}.
\end{eqnarray}
Here, for a field $f$, the left and right-handed couplings
are defined by  $g_L^f=I^f_3-e_f s_W^2$ and  $g_R^f=-e_f s_W^2$ with
$e_f$ being the electric charge in unit of $e$, 
and $I_3^f=\pm 1/2$ the corresponding third components of the weak isospin.
$f_c=2$ is a color factor.

The contribution of Fig.3 to $Zb_R\bar b_R$ coupling is found to be
\begin{eqnarray}\label{deltR2}
\Delta_R^{(\lambda'_{ij3})}&=&|\lambda'_{ij3}|^2 \frac{1}{16\pi^2}
\sum_{f,\tilde f} \left\{ B_1(M_b,M_f,M_{\tilde f})
+2\frac{g^{\tilde f}_L}{g^b_R}
C_{24}(p_b,-k,M_f,M_{\tilde f},M_{\tilde f})\right.\nonumber\\
& &  -\frac{g^f_L}{g^b_R}\left[0.5-2C_{24}-
M_Z^2(C_{11}-C_{12}+C_{21}-C_{23}) \right.\nonumber\\
& & \left. \left.+\frac{g^f_R}{g^f_L}M_f^2 C_0
\right] (k,-p_{\bar b},M_f,M_f,M_{\tilde f})\right\},
\end{eqnarray}
where the sum is performed over   
\begin{eqnarray}
(f,\tilde f)=\left \{ \begin{array}{l}
             (d^j,\tilde \nu^i_L)\\  (\nu^i,\tilde d^j_L)\\
             (u^j,\tilde e^i_L)\\  (e^i,\tilde u^j_L)
             \end{array}\right.
\end{eqnarray}

The contribution of Fig.4 to $Zb_L\bar b_L$ coupling 
is given by
\begin{eqnarray}\label{deltL}
\Delta_L^{(\lambda'_{i3k})}&=&|\lambda'_{i3k}|^2 \frac{1}{16\pi^2}
\left\{ B_1(M_b,M_{d^k},M_{\tilde \nu^i})+
 B_1(M_b,0,M_{\tilde d^k_R})\right.\nonumber\\
& & -2\frac{g^{\tilde \nu^i}_L}{g^b_L}
C_{24}(-p_b,k,M_{d^k},M_{\tilde \nu^i},M_{\tilde \nu^i})
 +2\frac{g^{\tilde d^k}_R}{g^b_L}
C_{24}(p_b,-k,0,M_{\tilde d^k_R},M_{\tilde d^k_R})\nonumber\\
& &  -\frac{g^{d^k}_R}{g^b_L}\left[0.5-2C_{24}-
M_Z^2(C_{11}-C_{12}+C_{21}-C_{23})\right]
(-k,p_{\bar b},M_{d^k},M_{d^k},M_{\tilde \nu^i})\nonumber\\
& & \left. +\frac{g^{\nu^i}_L}{g^b_L}\left[0.5-2C_{24}-
M_Z^2(C_{11}-C_{12}+C_{21}-C_{23})\right]
(k,-p_{\bar b},0,0,M_{\tilde d^k_R})\right\}.
\end{eqnarray}

\begin{table}
\caption{ The $1\sigma$ upper limits on $B$-violating 
top quark couplings $\lambda''_{3j3}$ from $R_b$
for scenario A ($M=250$ GeV, $\mu=-100$ GeV) and scenario B 
($M=100$ GeV, $\mu=-250$ GeV). The  $2\sigma$ upper limits 
are weaker than the perturbative unitarity bound of 1.25 and
thus not listed here.}
\label{table_1}
\vspace{7mm}
\begin{center}
\begin{tabular}{ccccc}
    & & & & \\
  &\multicolumn{2}{c}{ Scenario A }&\multicolumn{2}{c}{ Scenario B }\\
\cline{2-3} \cline{4-5} 
sfermion mass (GeV)
  & $\tan\beta=2$ &  $\tan\beta=30$ & $\tan\beta=2$ &  $\tan\beta=30$ \\ \hline
  100 &  1.23 &  1.18 &  1.07&  0.87\\
  200 &  1.30 &  1.26 &  1.06&  0.82\\
  300 &  1.39 &  1.36 &  1.12&  0.86\\
  400 &  1.48 &  1.45 &  1.19&  0.93\\
  500 &  1.57 &  1.55 &  1.27&  1.02\\
  600 &  1.65 &  1.64 &  1.36&  1.12\\
  700 &  1.77 &  1.73 &  1.45&  1.23 \\
  800 &  1.72 &  1.89 &  1.54&  1.33 \\
  900 &  1.91 &  1.90 &  1.63&  1.43 \\
 1000 &  2.01 &  2.00 &  1.73&  1.54 \\
\end{tabular}
\end{center}
\end{table}
\begin{table}
\caption{ The $1\sigma$ ( $2\sigma$) upper limits on $B$-violating 
top quark couplings $\lambda''_{3j3}$ from $R_{\ell}$
for scenario A ($M=250$ GeV, $\mu=-100$ GeV) and scenario B 
($M=100$ GeV, $\mu=-250$ GeV). These bounds are also applied to
$\lambda''_{3jk}$ if all sfermions have the same mass.}
\label{table_2}
\vspace{7mm}

\begin{center}
\begin{tabular}{ccccc}
    & & & & \\
  &\multicolumn{2}{c}{Scenario A}&\multicolumn{2}{c}{Scenario B}\\
\cline{2-3} \cline{4-5} 
sfermion mass (GeV)
  & $\tan\beta=2$ &  $\tan\beta=30$ & $\tan\beta=2$ &  $\tan\beta=30$ \\ \hline
  100 &  0.75(1.14) &  0.72(1.12) &  0.65(1.08) &  0.53(1.01) \\
  200 &  0.79(1.30) &  0.77(1.29) &  0.65(1.22) &  0.51(1.15) \\
  300 &  0.85(1.49) &  0.83(1.48) &  0.69(1.40) &  0.54(1.34) \\
  400 &  0.91(1.68) &  0.89(1.67) &  0.74(1.59) &  0.59(1.53) \\
  500 &  0.96(1.86) &  0.95(1.86) &  0.79(1.78) &  0.65(1.72) \\
  600 &  1.02(2.04) &  1.01(2.04) &  0.85(1.96) &  0.71(1.91) \\
  700 &  1.09(2.24) &  1.07(2.23) &  0.91(2.15) &  0.78(2.10) \\
  800 &  1.07(2.38) &  1.17(2.42) &  0.97(2.33) &  0.84(2.29) \\
  900 &  1.19(2.57) &  1.18(2.57) &  1.02(2.50) &  0.91(2.46) \\
 1000 &  1.25(2.75) &  1.24(2.75) &  1.09(2.69) &  0.98(2.64) \\
\end{tabular}
\end{center}
\end{table}
\begin{table}
\caption{  The $1\sigma$ ( $2\sigma$) upper limits on $L$-violating 
top quark couplings $\lambda'_{13k}$ from $R_e$
for scenario A ($M=250$ GeV, $\mu=-100$ GeV) and scenario B 
($M=100$ GeV, $\mu=-250$ GeV).}
\label{table_3}
\vspace{7mm}
\begin{center}
\begin{tabular}{ccccc}
    & & & & \\
  &\multicolumn{2}{c}{Scenario A}&\multicolumn{2}{c}{Scenario B}\\
\cline{2-3} \cline{4-5} 
sfermion mass (GeV)
  & $\tan\beta=2$ &  $\tan\beta=30$ & $\tan\beta=2$ &  $\tan\beta=30$ \\ \hline
  100 &  0.72(0.89) &  0.72(0.90) &  0.73(0.91) &  0.75(0.92) \\
  200 &  0.87(1.08) &  0.87(1.08) &  0.89(1.10) &  0.91(1.11) \\
  300 &  1.04(1.29) &  1.05(1.29) &  1.06(1.30) &  1.08(1.32) \\
  400 &  1.21(1.49) &  1.22(1.50) &  1.23(1.51) &  1.25(1.52) \\
  500 &  1.38(1.70) &  1.38(1.70) &  1.40(1.71) &  1.41(1.72) \\
  600 &  1.54(1.90) &  1.55(1.90) &  1.56(1.91) &  1.57(1.92) \\
  700 &  1.70(2.09) &  1.71(2.09) &  1.72(2.10) &  1.73(2.11) \\
  800 &  1.87(2.29) &  1.86(2.28) &  1.88(2.30) &  1.89(2.30) \\
  900 &  2.02(2.47) &  2.02(2.47) &  2.03(2.49) &  2.04(2.49) \\
 1000 &  2.17(2.66) &  2.17(2.66) &  2.19(2.67) &  2.20(2.68) \\
\end{tabular}
\end{center}
\end{table}
\begin{table}
\caption{  The $1\sigma$ ( $2\sigma$) upper limits on $L$-violating 
top quark couplings $\lambda'_{23k}$ from $R_{\mu}$
for scenario A ($M=250$ GeV, $\mu=-100$ GeV) and scenario B 
($M=100$ GeV, $\mu=-250$ GeV).}
\label{table_4}
\vspace{7mm}
\begin{center}
\begin{tabular}{ccccc}
    & & & & \\
  &\multicolumn{2}{c}{Scenario A}&\multicolumn{2}{c}{Scenario B}\\
\cline{2-3} \cline{4-5} 
sfermion mass (GeV)
  & $\tan\beta=2$ &  $\tan\beta=30$ & $\tan\beta=2$ &  $\tan\beta=30$ \\ \hline
  100 &  0.58(0.74) &  0.58(0.74) &  0.60(0.75) &  0.62(0.77) \\
  200 &  0.71(0.89) &  0.71(0.90) &  0.73(0.91) &  0.75(0.93) \\
  300 &  0.86(1.07) &  0.86(1.07) &  0.88(1.09) &  0.90(1.11) \\
  400 &  1.00(1.25) &  1.00(1.25) &  1.02(1.27) &  1.04(1.28) \\
  500 &  1.14(1.42) &  1.14(1.42) &  1.16(1.44) &  1.18(1.45) \\
  600 &  1.28(1.58) &  1.28(1.59) &  1.30(1.60) &  1.31(1.61) \\
  700 &  1.41(1.75) &  1.41(1.75) &  1.43(1.77) &  1.45(1.78) \\
  800 &  1.55(1.92) &  1.54(1.91) &  1.57(1.93) &  1.58(1.94) \\
  900 &  1.68(2.07) &  1.68(2.07) &  1.70(2.09) &  1.71(2.09) \\
 1000 &  1.80(2.23) &  1.81(2.23) &  1.82(2.24) &  1.83(2.25) \\
\end{tabular}
\end{center}
\end{table}
\begin{table}
\caption{  The $1\sigma$ ( $2\sigma$) upper limits on $L$-violating 
top quark couplings $\lambda'_{33k}$ from $R_{\tau}$
for scenario A ($M=250$ GeV, $\mu=-100$ GeV) and scenario B 
($M=100$ GeV, $\mu=-250$ GeV).}
\label{table_5}
\vspace{7mm}
\begin{center}
\begin{tabular}{ccccc}
    & & & & \\
  &\multicolumn{2}{c}{Scenario A}&\multicolumn{2}{c}{Scenario B}\\
\cline{2-3} \cline{4-5} 
sfermion mass (GeV)
  & $\tan\beta=2$ &  $\tan\beta=30$ & $\tan\beta=2$ &  $\tan\beta=30$ \\ \hline
  100 &  0.16(0.54) &  0.18(0.55) &  0.22(0.56) &  0.27(0.58) \\
  200 &  0.26(0.67) &  0.27(0.67) &  0.32(0.69) &  0.36(0.71) \\
  300 &  0.35(0.81) &  0.36(0.81) &  0.41(0.83) &  0.44(0.85) \\
  400 &  0.43(0.94) &  0.44(0.95) &  0.49(0.97) &  0.52(0.99) \\
  500 &  0.51(1.08) &  0.51(1.08) &  0.56(1.10) &  0.59(1.12) \\
  600 &  0.58(1.21) &  0.59(1.21) &  0.63(1.23) &  0.66(1.25) \\
  700 &  0.65(1.34) &  0.66(1.34) &  0.70(1.36) &  0.73(1.37) \\
  800 &  0.74(1.47) &  0.72(1.46) &  0.77(1.49) &  0.79(1.50) \\
  900 &  0.80(1.59) &  0.80(1.59) &  0.83(1.61) &  0.86(1.62) \\
 1000 &  0.86(1.71) &  0.86(1.71) &  0.90(1.73) &  0.92(1.74) \\
\end{tabular}
\end{center}
\end{table}
\begin{table}
\caption{ A summary of updated current bounds on $R$-violating top quark 
Yukawa couplings
          for slepton mass and squark mass of 100 GeV.}
\label{table_6}
\vspace{7mm}
\begin{center}
\begin{tabular}{|l|l|l|}
~~Top Quark R-violating Couplings~~ & Limits & Sources \\\hline
~~~~$\lambda^{\prime}_{131}$ &0.035 & APV, $2\sigma$ \\ \hline
~~~~$\lambda^{\prime}_{132}$ &0.75(0.92)  & $R_e$ at LEP, $1\sigma$ ($2\sigma$) 
                                                          \\ \hline
~~~~$\lambda^{\prime}_{133}$ &0.0007& $\nu_e$-mass, $1\sigma$ \\ \hline
~~~~$\lambda^{\prime}_{231}$ &0.22& $\nu_{\mu}$ scatter, $2\sigma$ \\ \hline
~~~~$\lambda^{\prime}_{232}$,~$\lambda^{\prime}_{233}$
                         &0.62(0.77) & $R_{\mu}$ at LEP, $1\sigma$ ($2\sigma$)
                                                                \\ \hline
~~~~$\lambda^{\prime}_{331}$,~$\lambda^{\prime}_{332}$,$\lambda^{\prime}_{333}$
 &0.27(0.58) & $R_{\tau}$ at LEP, $1\sigma$ ($2\sigma$) \\ \hline \hline
~~~~$\lambda''_{312}$,~$\lambda''_{313}$,
    ~$\lambda''_{323}$ 
 &0.75(1.14)  & $R_{\ell}$ at LEP, $1\sigma$  ($2\sigma$)\\
\end{tabular}
\end{center}
\end{table}
\begin{figure}
\begin{center}
\psfig{figure=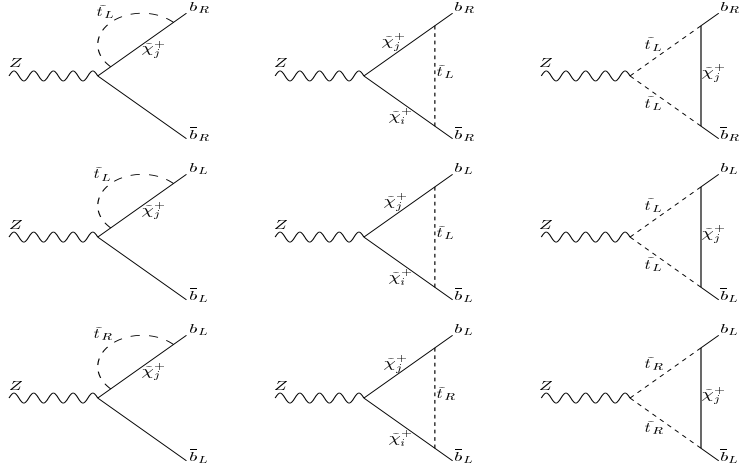,width=400pt,height=400pt,angle=0}
\end{center}
\vspace*{-5cm}
\caption{Feynman diagrams of chargino-stop loops
         which contribute to $Zb\bar b$ vertex.}
\end{figure}
\samepage
\begin{figure}
\begin{center}
\psfig{figure=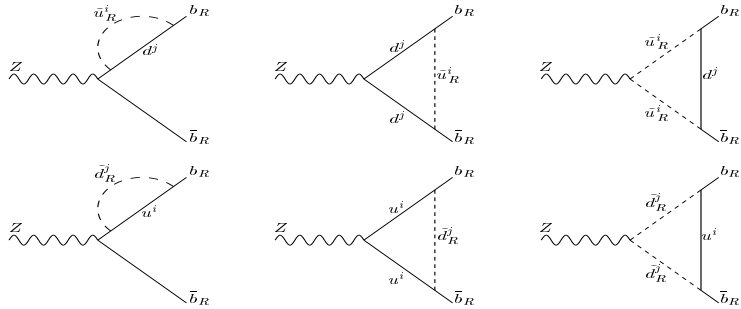,width=400pt,height=400pt,angle=0}
\end{center}
\vspace*{-7cm}
\caption{Feynman diagrams for the $B$-violating $\lambda''_{ij3}$ 
         contributions to $Zb_R\bar b_R$ vertex.
 $i$ and $j$ are flavor indices, with $i=1$,2 or 3 and $j=1$ or 2.}
\end{figure}
\eject
\begin{figure}
\begin{center}
\psfig{figure=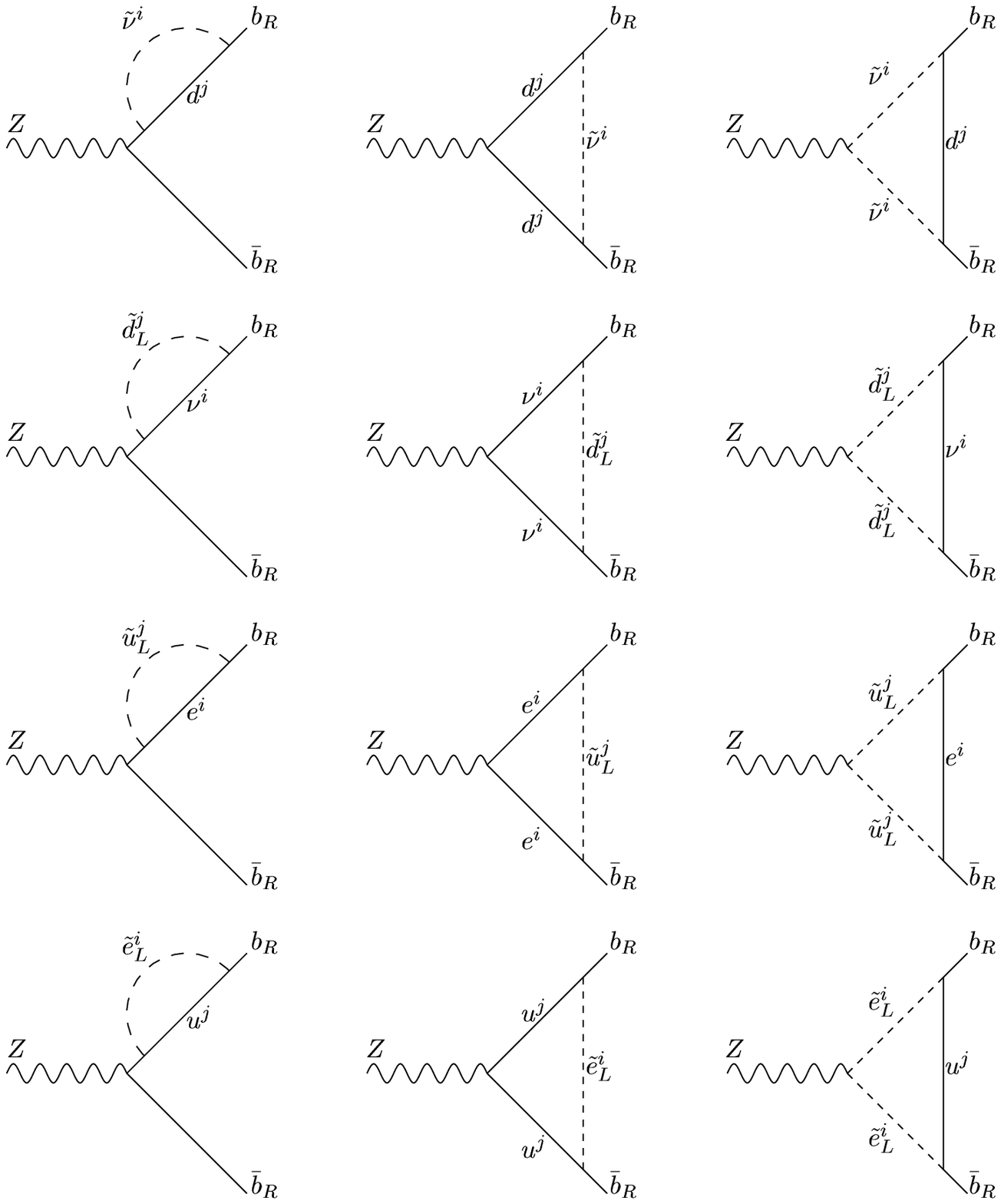,width=400pt,height=400pt,angle=0}
\end{center}
\vspace*{-3cm}
\caption{Feynman diagrams for the $L$-violating $\lambda'_{ij3}$ 
         contributions to $Zb_R\bar b_R$ vertex.
         $i$ and $j$ are flavor indices.}
\end{figure}
\samepage
\begin{figure}
\begin{center}
\psfig{figure=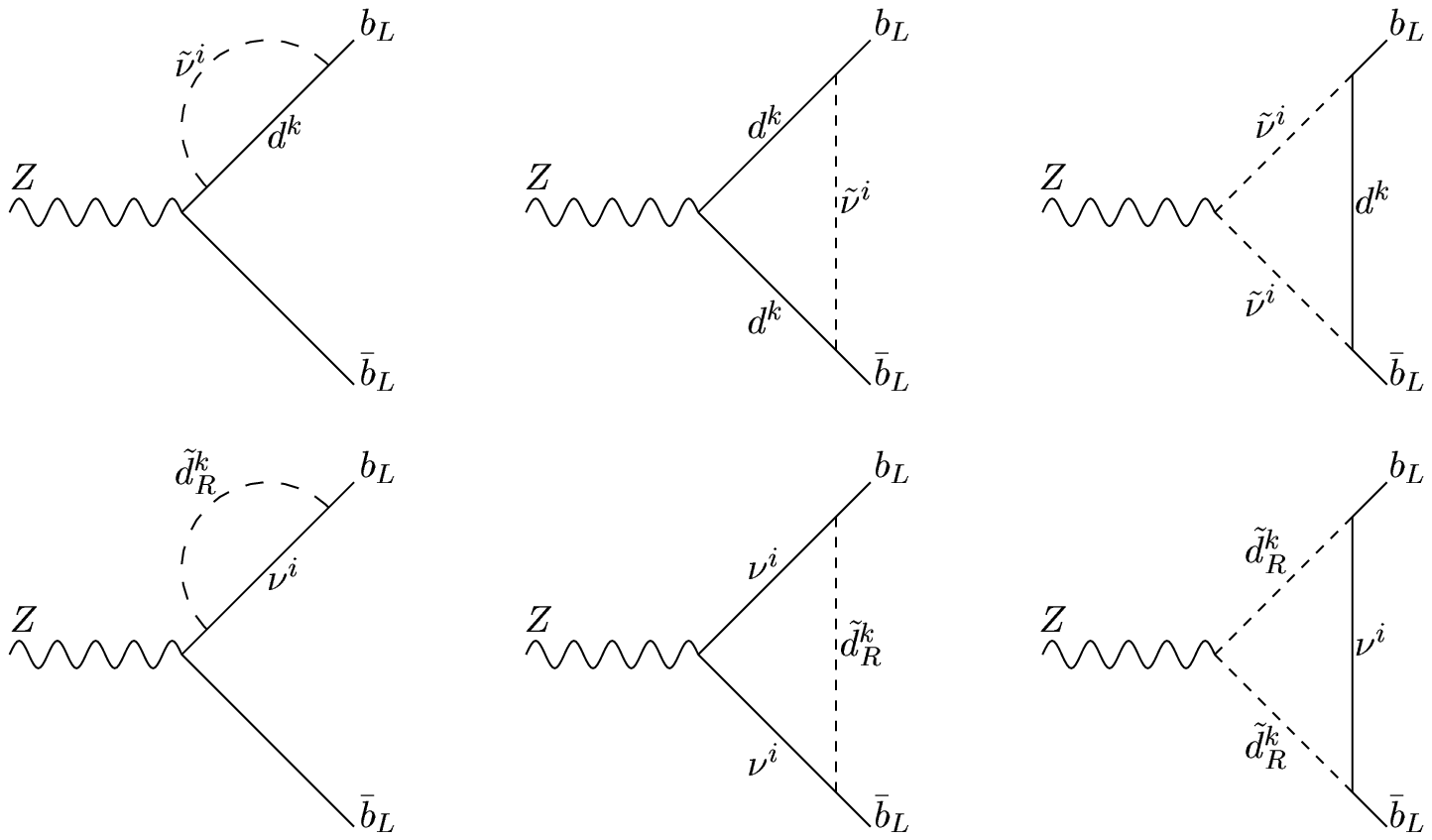,width=400pt,height=400pt,angle=0}
\end{center}
\vspace*{-7cm}
\caption{Feynman diagrams for the $L$-violating $\lambda'_{i3k}$ 
         contributions to $Zb_L\bar b_L$ vertex.
         $i$ and $k$ are flavor indices.}
\end{figure}
\begin{figure}
\begin{center}
\psfig{figure=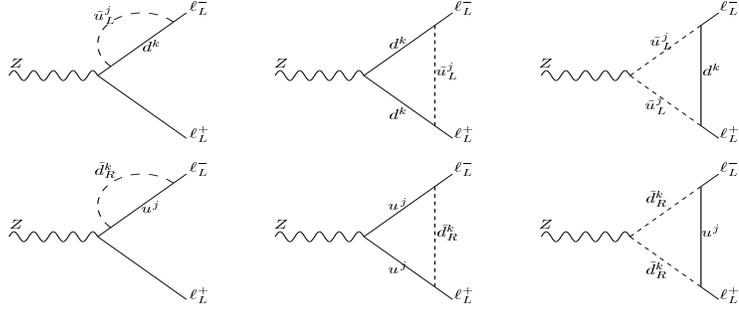,width=400pt,height=400pt,angle=0}
\end{center}
\vspace*{-6cm}
\caption{Feynman diagrams for the $L$-violating $\lambda'_{i3k}$ 
         contributions to left-handed $Z\ell_L \ell_L$ vertex.
         $i$ and $k$ are flavor indices. }
\vfil
\end{figure}
\newpage
\begin{figure}
\begin{center}
\psfig{figure=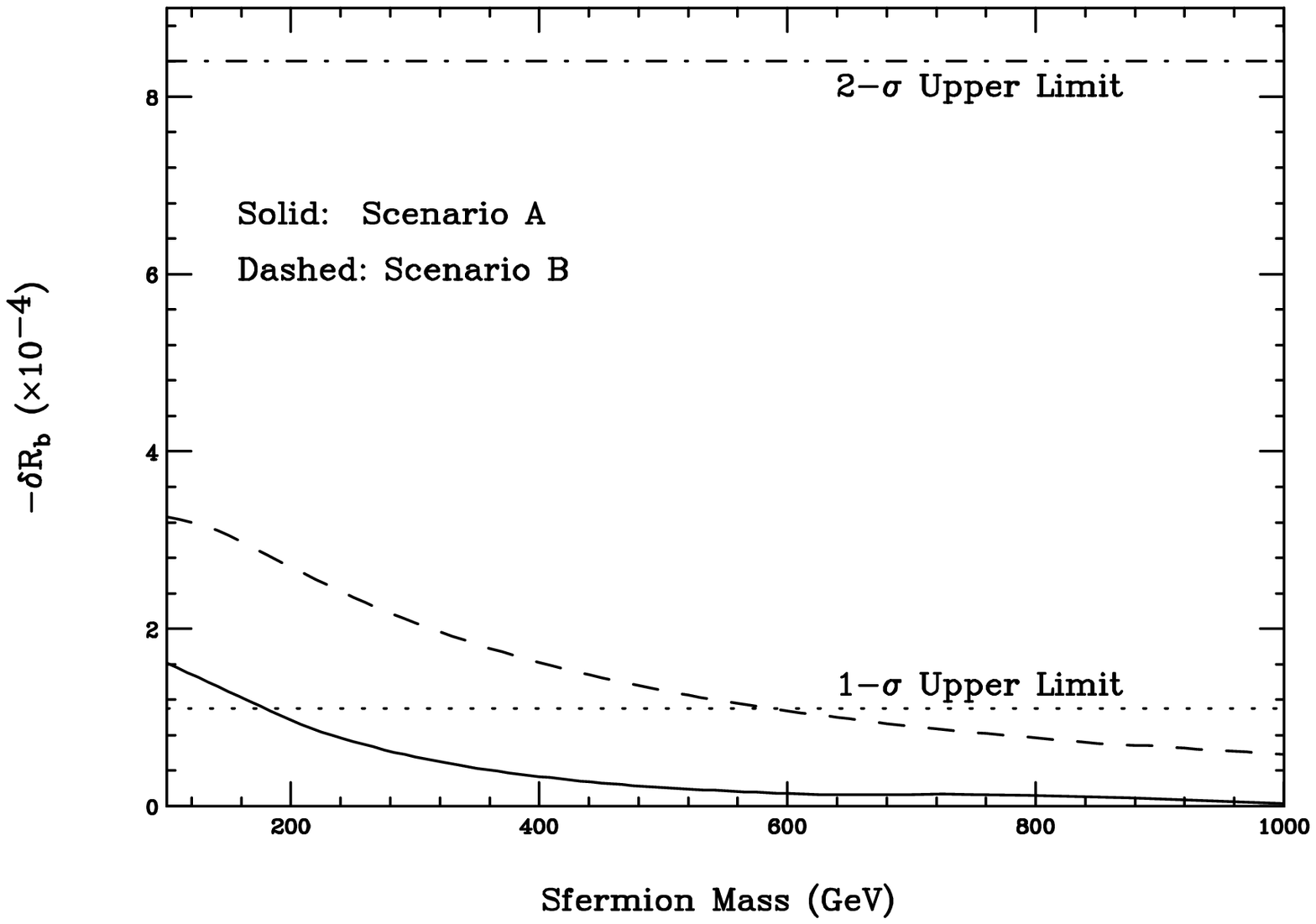,width=400pt,height=400pt,angle=90}
\end{center}
\caption{$\delta R_b$ in the presence of 
         $\lambda''_{3j3}=1.25$ versus sfermion mass 
    for scenario A ($M=250$ GeV, $\mu=-100$ GeV) and scenario B 
($M=100$ GeV, $\mu=-250$ GeV).}
\vfil
\end{figure}
\newpage
\begin{figure}
\begin{center}
\psfig{figure=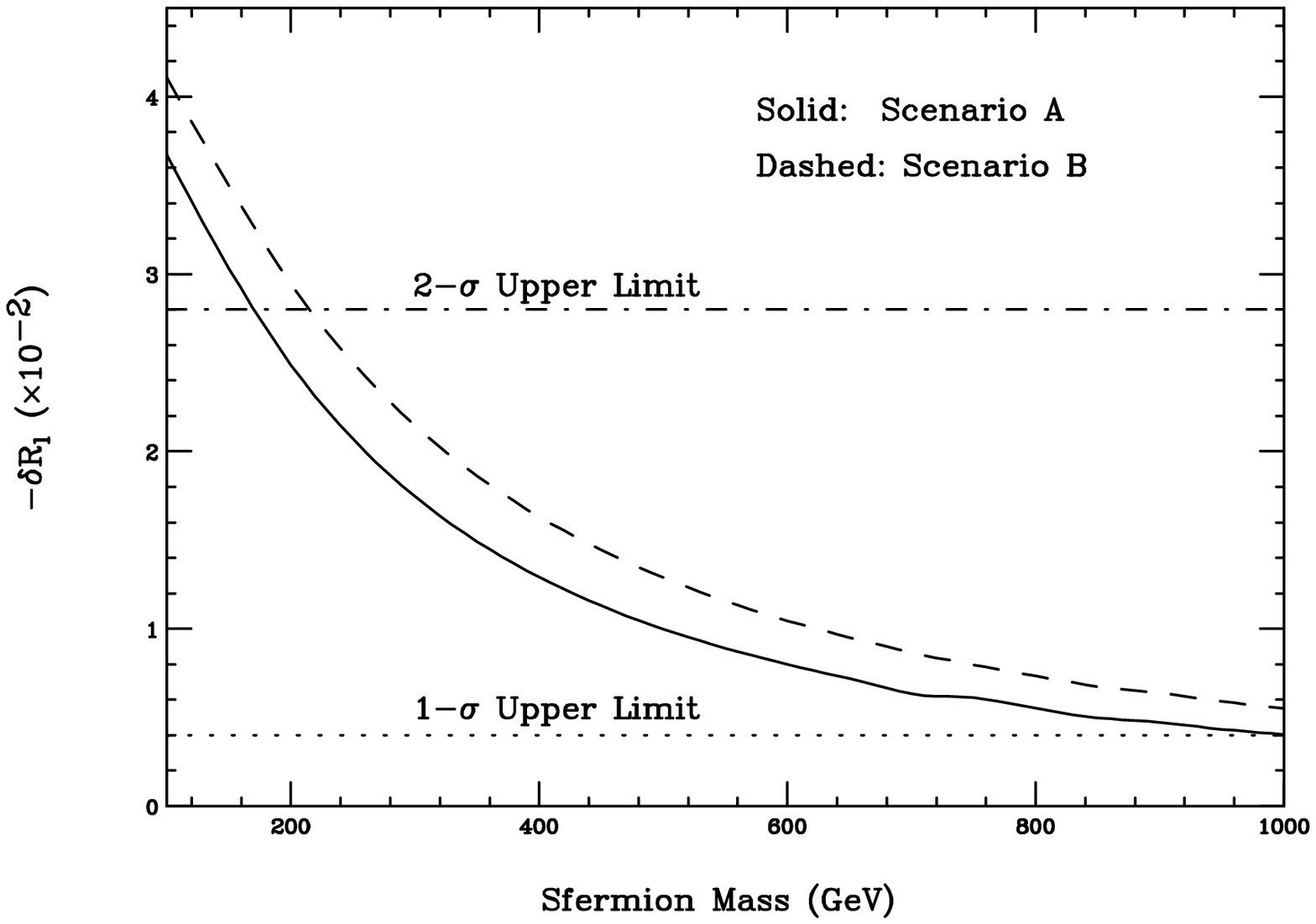,width=400pt,height=400pt,angle=90}
\end{center}
\caption{$\delta R_{\ell}$ in the presence of 
         $\lambda''_{3j3}=1.25$ versus sfermion mass 
    for scenario A ($M=250$ GeV, $\mu=-100$ GeV) and scenario B 
($M=100$ GeV, $\mu=-250$ GeV).}
\vfil
\end{figure}
\newpage
\begin{figure}
\begin{center}
\psfig{figure=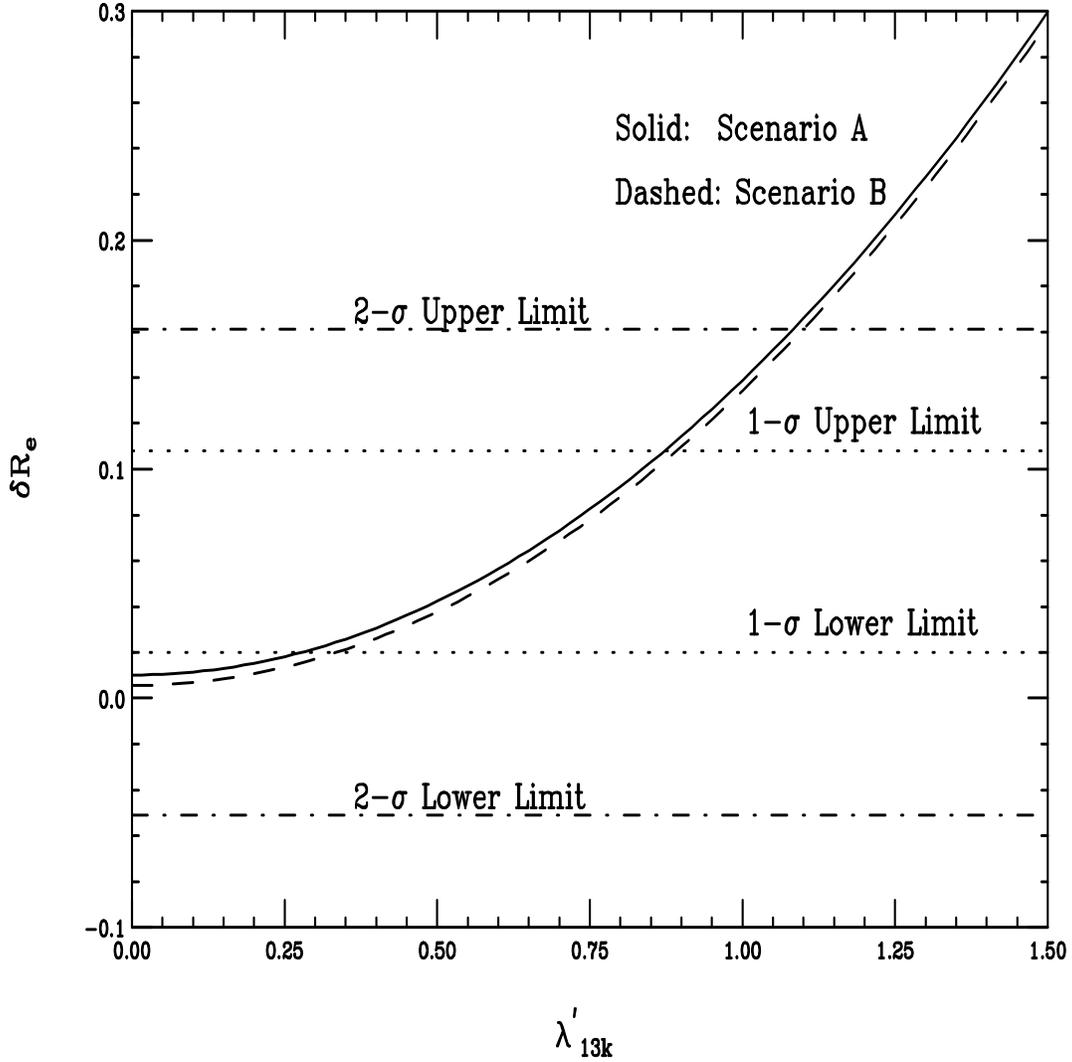,width=400pt,height=400pt,angle=90}
\end{center}
\caption{$\delta R_e$ versus $\lambda'_{13k}$ 
         for versus sfermion mass of 200 GeV, 
    under scenario A ($M=250$ GeV, $\mu=-100$ GeV) and scenario B 
($M=100$ GeV, $\mu=-250$ GeV).}
\vfil
\end{figure}
\newpage
\begin{figure}
\begin{center}
\psfig{figure=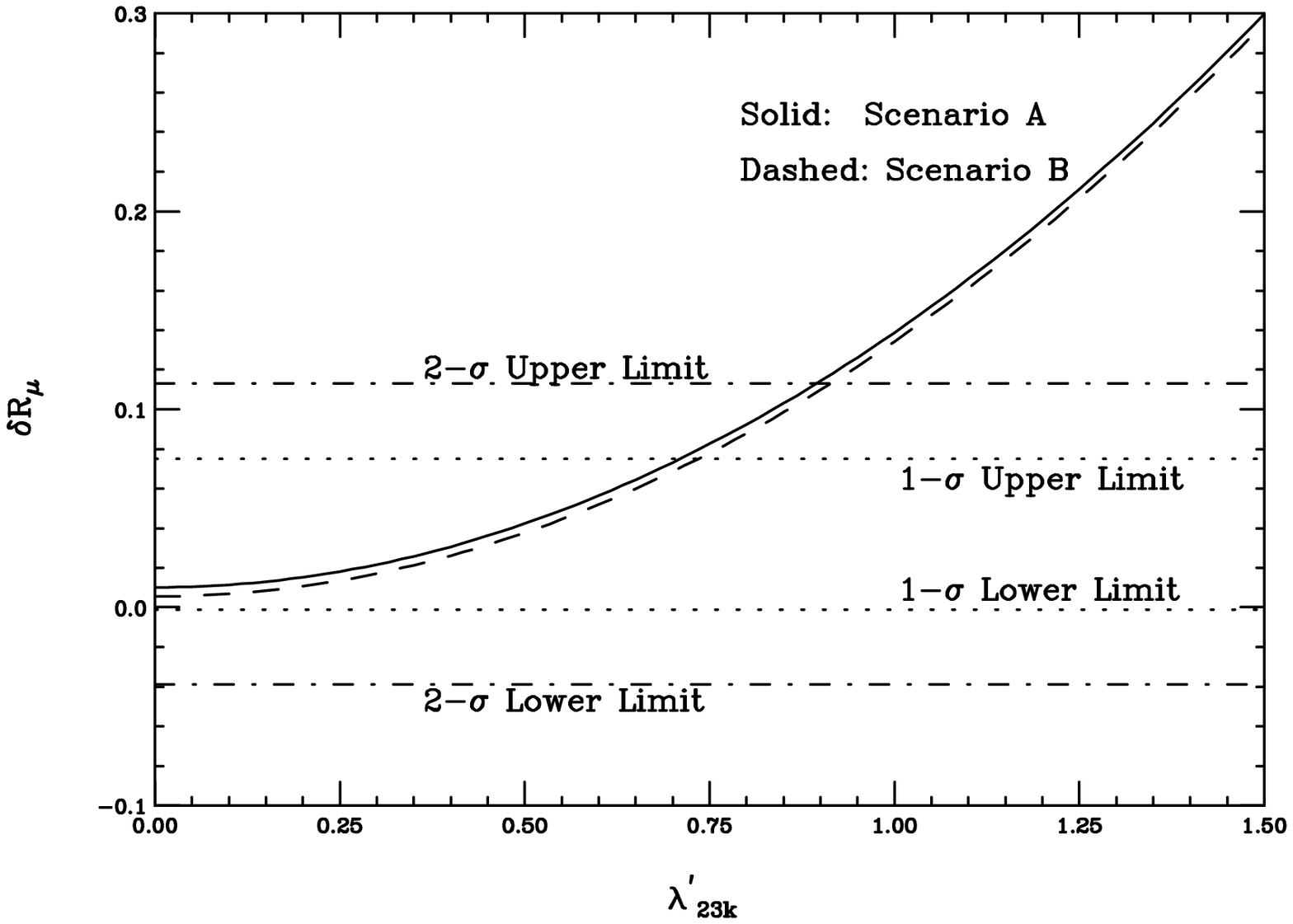,width=400pt,height=400pt,angle=90}
\end{center}
\caption{Same as Fig. 8, but for $\delta R_{\mu}$ versus $\lambda'_{23k}$.}
\vfil
\end{figure}
\newpage
\begin{figure}
\begin{center}
\psfig{figure=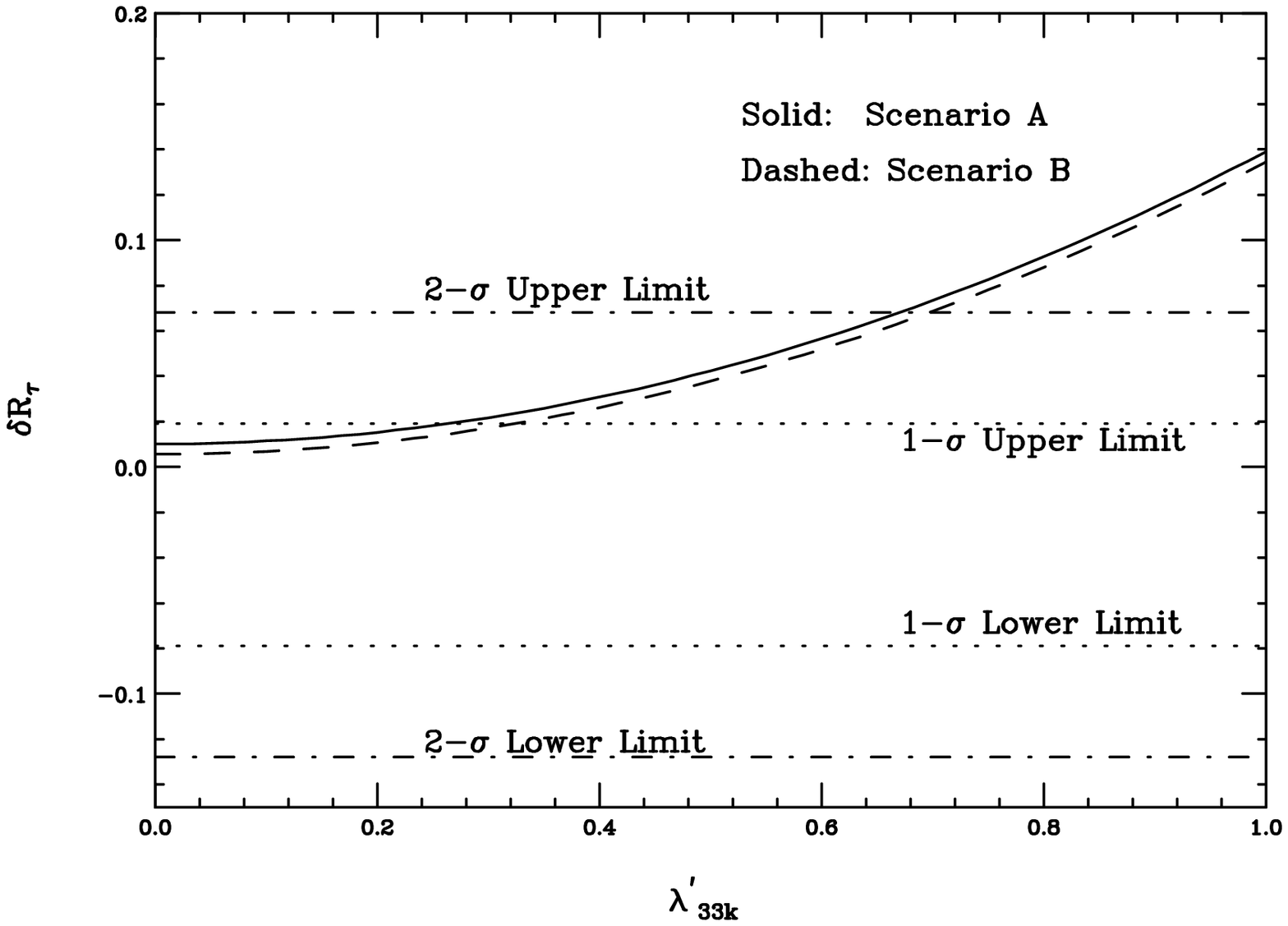,width=400pt,height=400pt,angle=90}
\end{center}
\caption{Same as Fig. 8, but for $\delta R_{\tau}$ versus $\lambda'_{33k}$.}
\vfil
\end{figure}
\end{document}